\documentclass[twocolumn]{aastex62}

\usepackage{ulem}
\usepackage{yfonts}
\usepackage{hyperref}
\usepackage{subfigure}
\received{July 9, 2018}
\revised{October 6, 2018}
\accepted{October 8, 2018}

\submitjournal{AAS Journals}

\shorttitle{Connecting \emph{Kepler} Starspots and Flares }
\shortauthors{Roettenbacher \& Vida}

\begin{document}

\title{The connection between starspots and flares on main-sequence \emph{Kepler} stars}

\correspondingauthor{Rachael M.\ Roettenbacher}
\email{rachael.roettenbacher@yale.edu}

\author[0000-0002-9288-3482]{Rachael M.\ Roettenbacher}
\affiliation{Department of Astronomy\\
Stockholm University \\
SE-106 91 Stockholm, Sweden}
\affiliation{Yale Center for Astronomy and Astrophysics\\
Department of Physics\\
Yale University\\
New Haven, CT 06520 USA}

\author[0000-0002-6471-8607]{Kriszti\'an Vida}
\affiliation{Konkoly Observatory\\
MTA CSFK\\
H-1121 Budapest, Konkoly Thege M.\ \'ut 15-17 Hungary}

\begin{abstract}

Starspots and flares are indicators of stellar magnetic activity that can both be studied in greater detail by utilizing the long-term, space-based archive of the \emph{Kepler} satellite.  
Here, we aim to investigate a subset of the \emph{Kepler} archive to reveal a connection between the starspots and the stellar flares, in order to provide insight into the overall stellar magnetic field.
We use the flare-finding algorithm FLATW'RM in conjunction with a new suite of algorithms that aim to locate the local minima caused by starspot groups.  We compare the phase difference between the time of maximum flux of a flare and the time of minimum stellar flux due to a starspot group.  
The strongest flares do not appear to be correlated to the largest starspot group present, but are also not uniformly distributed in phase with respect to the starspot group.  The weaker flares, however, do show an increased occurrence close to the starspot groups.

\end{abstract}


\section{Introduction} \label{sec:intro}

Stellar activity can manifest as a number of magnetic phenomena, most prominently as starspots and flares.  In fully convective stars or cool stars with convective outer envelopes, starspots are dark regions where the local magnetic field suppresses the convection \citep[][and references therein]{ber05,str09}.  Flares are the energetic results of magnetic field line reconnection during which there are rapid increases in the stellar flux due to sudden energy release \citep[particularly in the optical and UV; see][and references therein]{pri02}.

The \emph{Kepler} space telescope launched in 2009 with the primary goal of detecting Earth-like planets in Earth-like orbits around Sun-like stars \citep{bor10,koc10}, but the observations have provided a wealth of data for stellar activity studies.  For example, \citet{bas10} provided an early look at \emph{Kepler} activity, while \citet{wal11,dav14} focused on the stellar flares.  These studies are of particular interest to improving our understanding of stellar activity, as many of the observed stars behave significantly differently than the Sun with larger starspots and with more frequent, as well as stronger, flares \citep[e.g.,][]{mae12,roe13,vid16}.  Beyond stellar astrophysics, understanding stellar activity in detail will reveal vital information for the interaction between stars and their planetary systems \citep[e.g.,][]{vid17,roe17a}.

On the Sun, spot groups can be resolved into individual sunspots, but this is not possible for other stars, even with interferometry where a small number of spotted stars can be resolved \citep[e.g.,][]{roe16b,roe17b}.  Because of this, the starspots that we see evidence of photometrically are likely to be analogous to sunspot groups.  Additionally, on the Sun, the origin of solar flares can be determined, but the limitations of stellar photometry prevent the position of the flare on the stellar disk from being determined.  
Here, we investigate whether the stellar flares show a connection using the location of the starspot groups with a collection of stars showing evidence of both starspots and flares.  In Section \ref{sec:obs}, we describe the observations used for this study.  In Section \ref{sec:analysis}, we describe our method of analysis for the light curves, and in Section \ref{sec:discussion}, we describe our results.  In the final section, Section \ref{sec:conclusions}, we conclude.  In the Appendix, we include a discussion of the model light curves we used to test our analysis method and those results; a simple model of a spotted, flaring star with flares connected with the spots; and further discussion of subsets of the results.

\section{Observations} \label{sec:obs}

The \emph{Kepler} space telescope observed $\sim 150,000$ stars during the  its four years of observation \citep{bor10,koc10}.  Among those stars were 34,030 main-sequence stars for which \citet{mcq14} found rotation periods based on rotational modulation using an autocorrelation method.  \citet{dav16} analyzed the flares of the \emph{Kepler} catalog, finding 4,041 flaring dwarf stars (stars with 100 or more potential flaring events with at least ten of those events above a completeness threshold).  The intersection of these two works is a set of 402 stars, which we use as our initial sample.  
\citet{dav16} uses both long (30-minute exposure) and short (1-minute exposure) cadence data, but here we use only the long cadence data in order to keep our sample homogeneous and because the timescale of stellar rotation is on the order of days so the short-cadence light curves provide no additional information in our science case, as the rotational timescale is much longer than the uncertainties caused by light curve sampling.  
The stars in this sample are all assumed to be on the main sequence \citep[following][]{mcq14} and range from late-F to mid-M spectral types \citep[based on masses from][]{dav16}.

\section{Analysis} \label{sec:analysis}
This subset of \emph{Kepler} targets represent some of the most active stars in the archive.  Here, we analyze the potential correlation between the position of starspots and the occurrence of flares.
We are developing a suite of codes for analyzing stellar activity, particularly with respect to understanding activity in the context of potential planetary hosts, called Stellar Activity for Understanding and Characterizing Exoplanetary Systems (SAUCES)\footnote{SAUCES is currently written in \texttt{IDL} and is available at \url{https://github.com/RMRoettenbacher/SAUCES}.  This suite of codes will be expanded with future studies to further investigate stellar activity.  \textbf{Codes will be available upon publication.}}.

Here, SAUCES uses the long-cadence \emph{Kepler} light curve\footnote{The \emph{Kepler} light curves have been obtained from the Barbara A.\ Mikulski Archive for Space Telescopes (MAST).} \citep{tho16}.  The  Simple Aperture Photometry (SAP) long-cadence light curve is the flux in the custom apertures created for the star for 270 integrations spanning 29.424 minutes after the Photometric Analysis module of the \emph{Kepler} Science Operations Center data-processing pipeline has been applied.  The Pre-search Data Conditioning (PDC) light curve has had an additional PDC module applied to remove astrophysical signatures in order to better isolate transits and eclipses.  The PDC module can eliminate the signatures of starspots, so we use the SAP light curve.  For SAUCES, individual data points are removed if either the SAP or PDC flux values are negative or if any of the SAP quality flags are listed. 
To remove the systematics inherent in the \emph{Kepler} data while preserving stellar astrophysics, SAUCES applied cotrending basis vectors (CBVs) with code adapted from \texttt{kepcotrend.pro}\footnote{Currently available at \url{http://www.lpl.arizona.edu/~ bjackson/idl\_code/kepcotrend.pro}.}, developed by B.\ Jackson and N.\ Thom.  The SAP light curves are median-divided to account for the flux jumps across the different quarters of observation.

SAUCES then performs a period search on the CBV-removed, median-divided light curves using a Fast Fourier Transform (FFT) to detect the period.  We recognize that there are other period-finding algorithms available \citep[e.g., autocorrelation;][]{mcq14}, but the FFT method lends itself well to understanding if a rotation period is indicative of a spotted surface or a more regular light curve, such as those from pulsations or eclipsing binaries.  Because pulsations and eclipsing binaries are periodic with minimal variation from cycle to cycle, the periodogram of such a signal will resemble a Dirac-$\delta$ function at the periods evident in the data.  Conversely, starspots are potentially not as regular---potentially exhibiting evolution (growth or decay) and drift on the surface, on timescales as short as a rotation period.  Differential rotation could also cause a periodogram signature that is more complex than Dirac-$\delta$ functions \citep[e.g.,][]{ola03,rei13}.  While active longitudes are possible \citep[e.g.,][]{ola06,roe16a}, starspots often form at different latitudes across the stellar surface \citep[e.g.,][]{roe13,vid14,oza18}.  Because starspots vary over time (shape, size, and potentially even latitude), the signatures of longer observations can yield an ensemble of peaks around the strongest frequency.   If a star has more than one spot at different latitudes evolving over the  length of observation, those multiple starspots at different latitudes will result in periodogram peaks at different periods due to differential rotation \citep[e.g., Figure 8 of][]{roe16a}. 

By running a FFT on the \emph{Kepler} light curves of the stars at the intersection of the \citet{mcq14} and \citet{dav16} catalogs, SAUCES can distinguish between stars that belong in the sample of flaring, spotted stars and those that have mistakenly received the classification (e.g., RR Lyrae variables, non-radially pulsating stars, eclipsing binaries) based on the characteristics of the periodogram. Of the 402 stars in the cross-section of the \citet{mcq14} and \citet{dav16} samples, SAUCES recategorizes 227 stars as eclipsing binaries or pulsating stars.  Additionally, we categorized the light curves as likely starspots or not based on a by-eye identification aimed to eliminate stars with light curves that suggest they are $\gamma$ Doradus or $\delta$ Scuti stars (for examples of these light curves and a detailed discussion, see \citealt{uyt11}).  
Here, we will focus only on the 119 stars that both of these methods have designated as spotted. We acknowledge that this is dramatically different from the initial 402 stars in the cross-section of the \citet{mcq14} and the \citet{dav16} catalogs, but our aims here require a sample of  light curves that show evidence of evolving starspots. 

\begin{figure*}
\begin{center}
\includegraphics[scale=0.6,angle=90]{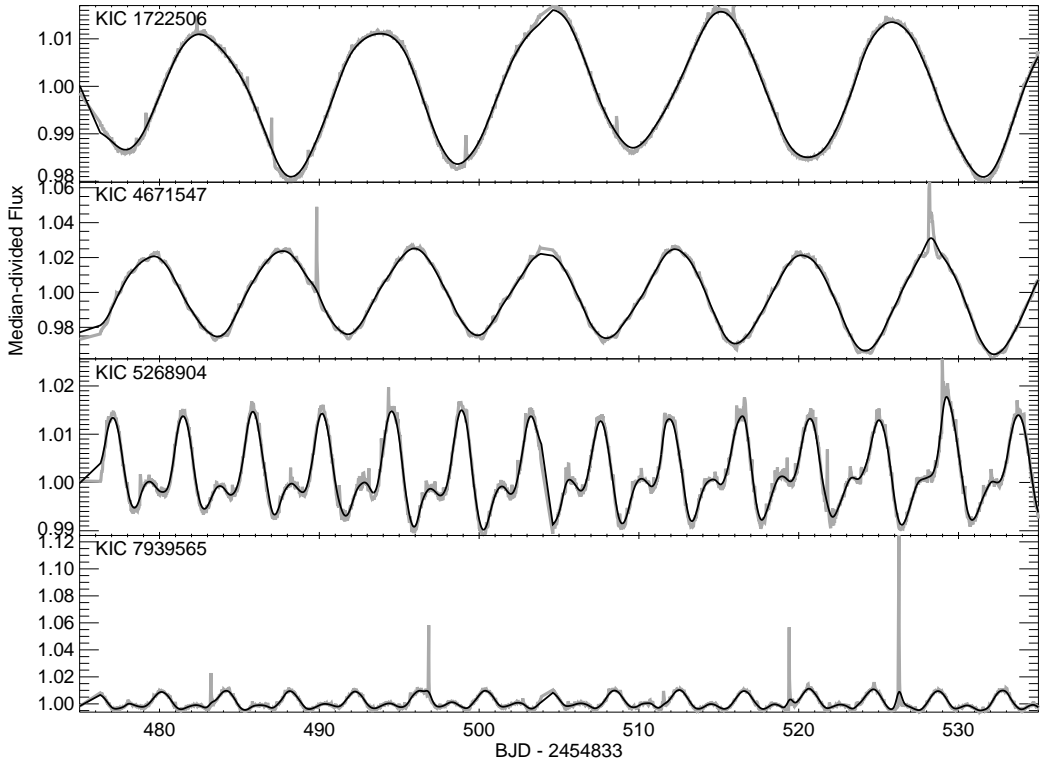}
\end{center}
\caption{Sample \emph{Kepler} long-cadence light curves (in gray) for four spotted, flaring stars in our sample. Overplotted (in black) are the light curves smoothed with a Gaussian kernel of width $0.1 \times P_\mathrm{rot}$.  These smoothed light curves are then used to determine the time of local minima for locating the position of starspots facing \emph{Kepler}.  Near several of the minima caused by starspots shown, flares occur, distorting the shape of the smoothed light curve.  This effect is accounted for by SAUCES, as described in Section \ref{sec:analysis}. }
\label{lcs}
\end{figure*}

SAUCES then locates the local minima in the light curves created by the rotational modulation of starspot groups.  The light curve is first smoothed with a Gaussian kernel with a width of $0.1 \times P_\mathrm{rot}$ to remove noise that could create false local minima, where $P_\mathrm{rot}$ is the rotation period estimated by the FFT method mentioned above (with the associated \citet{mcq14} period used as a starting parameter for centering the FFT search to reduce computation time; see Figure \ref{lcs}).  Once the light curve is smoothed, local minima are recorded simply by finding the lowest point in moving windows of about 10.5 hours, which was chosen to allow SAUCES to find multiple minima in a starspot group should they be present.  

The CBV-corrected light curves are separately run through our flare-finding algorithm FLAre deTection With Ransac Method \citep[FLATW'RM\footnote{FLATW'RM is available at \url{https://github.com/vidakris/flatwrm}.};][]{vid18}.  
FLATW'RM uses a machine-learning algorithm to give a robust model of the light curves in order to detect flare events and uses a voting system implemented to keep false positive detections to a minimum. 
For further details on the FLATW'RM algorithm, see \citet{vid18}. 
FLATW'RM detects flares and reports the times the flare starts and ends, the time of maximum flux, and the maximum percent increase of flux over the light curve around the flare, along with estimated flare energies, either from the raw light curve or by fitting an analytic flare model.  Here, we require the events to consist of at least three sequential data points in order to be detected and analyzed by FLATW'RM.  We note that the flares found by FLATW'RM (with no restrictions on flux increases) are fewer in number than \citet{dav16} reported, and, in many cases, significantly fewer.  There are, however, seven stars for which FLATW'RM identified more events as flares than \citet{dav14}.  The discrepancy is potentially due to different spurious events that fit the selection criteria, but are not flares.  We account for this discrepancy by focusing our analysis to potential flares with flux increases $\ge 1\%$, and \citet{dav16} identifies flares that are above a $68\%$ completeness threshold. 

SAUCES compares the time of the observed flare maximum and the time of the starspot minimum, where the flux has decreased the most due to the presence of a starspot group.  We compute the difference in phase of flare maximum and the phase of starspot minimum, $\Delta \varphi$.   Strong flares occurring in or near the center of the light curve minimum will distort the shape of the light curve and will affect the smoothed light curve, shifting the location of the local minimum.  In the event that the time of flare maximum is less than 0.40 days from the flare minimum,
we recalculate the time of the light curve local minimum by shifting one rotation period before and one rotation period after and locate the local minima there.  We then use the average of those two local minima times to assign as the time of the local minimum nearest to the flare.  In doing so, we assume that the starspot is somewhat stable over three rotation periods \citep[some M dwarfs e.g.,][show daily, but insubstantial variations in starspots]{vid16} and any changes in the location of the spot in phase is due to differential rotation (a slightly different period for the spot than that assigned to the star) and not due to a drift in latitude or dramatic evolution.  If a flare is also located in one of the neighboring rotations, SAUCES readjusts the time of local minimum using the next spot minima (two rotations before the flare event and two rotations after).  If there is a flare in those one of those minima, SAUCES does not continue searching further away because many of the stars in the \emph{Kepler} sample show dramatic evolution on timescales shorter than five rotations.  In these cases, SAUCES assigns this last calculation for $\Delta \varphi$.  For our sample, this occurs for $1.8\%$ of the flares observed and will not significantly affect our results.  
If local minima are reported near a gap in the data ($\ge 0.5$ days), we also apply these same shifts using the surrounding minima to determine $\Delta \varphi$.  

It may be the case that a flare starts and peaks more than 0.40 days away from the nearest local minimum, but the flare is of sufficiently high energy that it will affect the light curve for a long enough time after the initial brightening that the tail will significantly alter the shape of the light curve nearby.  For flares that end within 0.40 days of the local minimum, we apply the same readjustments to the location of the local minimum as described above.

The strength of a flare dictates the lifetime of the flare, and, therefore, how long it affects the light curve.  In the event of an isolated, single flare, predicting this lifetime and the flare's impact on the light curve could be done through the models determined by FLATW'RM.  However, we use the same, conservative length of impact for all the flares to account for the possibility that the flares are compound events leading to an incorrect model from FLATW'RM, as the relatively sparse sampling of the long-cadence \emph{Kepler} light curves and the exponential nature of the events can yield potentially unreasonable fits.

Additionally, we only look at the flares that occur when only one resolved starspot group is present on the stellar surface during the rotation in which the flare occurs.  With this restriction, we can clearly identify the flares that occur when the single starspot group is present ($-0.25 < \Delta \varphi < 0.25$) or when it is on the opposite side of the star ($\Delta \varphi > 0.25$ and $\Delta \varphi < -0.25$).  We also eliminate flares that have start times within two data points of a gap in the data ($\ge 0.5$ days) or end times within one data point of a gap in the data to further eliminate spurious results.  For examples and tests of this method, see Appendices A and  B.

\section{Discussion} \label{sec:discussion}

\begin{figure}
\begin{center}
\includegraphics[scale=0.35,angle=90]{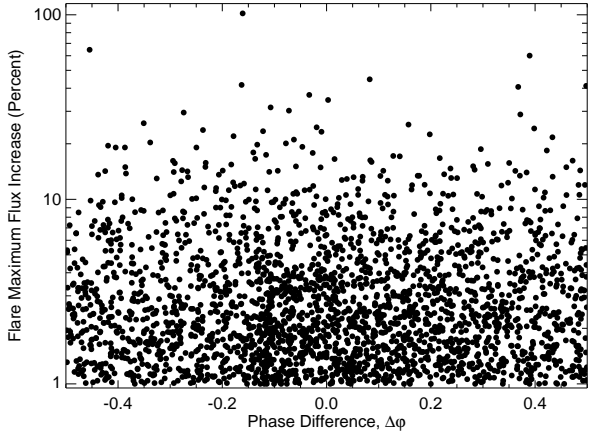}
\end{center}
\caption{Flare strengths as a function of phase difference, $\Delta \varphi$, for the 111 flaring, spotted stars in our sample. Each black dot represents one flare.  All flares with flux increases of $1\%$ or more are plotted.  Note that the ordinate axis is presented on a logarithmic scale.  }
\label{strengths}
\end{figure}

\begin{figure}
\begin{center}
\includegraphics[scale=0.35,angle=90]{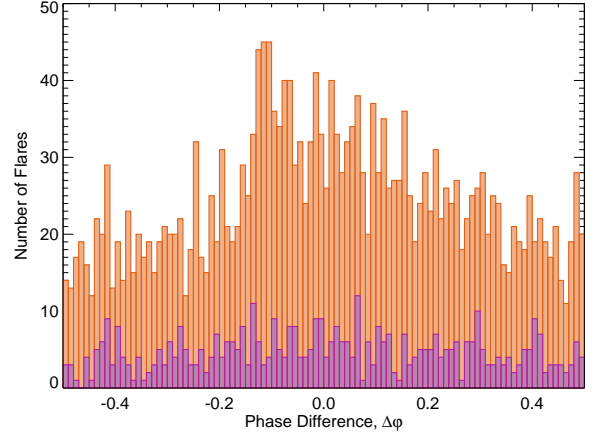}
\end{center}
\caption{Histogram of the number of flares occurring in bins of phase difference $\Delta \varphi = 0.01$.  The orange bins represents the flares that have flux increases between $1-5\%$, and the purple bins represent the flares that have flux increases $>5\%$.   Together, the purple and orange regions represent the total number of flares with flux increases $\ge 1\%$ that occur in each phase bin for the 111 flaring, spotted stars in our sample.}
\label{hist}
\end{figure}

For the sample of 119 stars, FLATW'RM detected 2447 flares that increase the flux of the star by $1\%$ or more in the light curves of 111 of the stars.  We plot the strength of each flare plotted against the phase difference, $\Delta \varphi$, in Figures \ref{strengths}.  In Figure \ref{hist}, we plot the number of flares in phase difference bins of $\Delta \varphi = 0.01$ for flares with an increase of at least $1\%$ in flux (with flux increases between $1-5\%$ and $>5\%$, inclusive, for a total of 1972 and 475 flares, respectively).   On solar-type stars, flares with increases in brightness of $0.1-1\%$ are considered superflares \citep[e.g.,][]{mae12}.  As all of the flares observed increase the overall stellar flux in the \emph{Kepler} bandpass by at least $1\%$, the flares in this sample are dramatically stronger than solar flares across the observed spectral types, with the strongest flares occurring in the lower mass stars. 

We performed a one-sided Kuiper test (an invariant Kolmogorov--Smirnov test appropriate for circular variables)
to determine if the phase differences between the starspot minima and the maximum amplitude of the flares were drawn from a uniform sample, which is expected after the results of \citet{hun12, roe13}; \citet{doy18}, and others for  active stars showing evidence that the  flares were possibly uniformly distributed.  
Differing from previous studies, this  work combines the results from 111 stars indicating that the phase differences between the starspot minima and the flare maxima are not drawn from a uniform distribution ($p \approx 0$ for flares with flux increases $> 1\%$ and $p = 0.002$ for flares with flux increases $>5 \%$).  In Figure \ref{hist}, there is a  distinct peak in the distribution centered around a zero phase difference between the starspot minimum and the flare maximum  (note that the starspot groups are in the field of view between $-0.25 \lesssim \Delta \varphi \lesssim 0.25$).  This suggests that the flares may be correlated with the starspots such that the magnetic fields causing the large starspot group are more frequently reconnecting to make flares than the magnetic fields elsewhere on the star.
Interestingly, when we look only at the strong flares (only the purple distribution in Figure \ref{hist}), the peak is not nearly as prominent, suggesting that the small flares are more tied to the location of the starspots than the large flares.

We considered whether this trend would change dependent upon spectral type.  In order to investigate this, we divided the sample at approximately the boundary between G and K stars, around $0.78 M_\odot$ \citep{dav16}.  This gave a sample of 57 F/G stars and 54 K/M stars.  While the K and M stars were more active, more weak flares occur close to the starspot in both the subsample of F/G stars and the subsample of K/M stars (see Appendix C).

Long-term monitoring of the fully convective M-dwarf V374\,Peg (M4V, $P_\mathrm{rot} \approx 0.44$ days) has shown that flares can occur anywhere on the surface and are not necessarily associated with large starspot groups \citep{vid16}, although on that object, stronger flares seemed to be associated with one of the active regions. This is in agreement with the numerical model of \cite{yad15} that suggests that magnetic field is distributed over the stellar surface in these type of stars, implying a flare distribution independent of rotation phase. Additionally, \citet{haw14} and \citet{lur15} investigated the relationship between starspots and flares on the M dwarfs GJ 1243 (M4, $P_\mathrm{rot} = 0.5927 \pm 0.0002$ days), GJ 1245A (M5, $P_\mathrm{rot} = 0.2632 \pm 0.0001$ days), and GJ 1245B (M5, $P_\mathrm{rot} = 0.709 \pm 0.001$ days).  In all three cases, they found that there was no correlation between starspot location and flare occurrence.  They suggest that the stars may have polar spots that do not cause periodic changes in the light curve but are associated with flares.  Interestingly, the light curves of V374 Peg, GJ 1243, and GJ 1245AB are all varying (evolving) very slowly, appearing nearly constant over the years of observation.  This is a distinct difference from the stars in our sample, which show surface evolution.

The shorter term study of 34 dwarfs in a field of \emph{K2} by \citet{doy18} reported no evidence of correlation between starspot location and flare timing.  While the goal was similar, their sample was significantly different from ours:  short-cadence \emph{K2} light curves are used, allowing for the detection of shorter, less-energetic flares; the lengths of observation ($\sim 80$ days) are significantly shorter than for the \emph{Kepler} light curves; $29\%$ would be rejected by our analysis, as they have rotation periods below 1 day, and they were unable to determine the periods of another $48\%$ of their sample given their short light curves; and several of their targets exhibit more than one spot structure at a time.  Of the seven of their stars with confirmed periods over $1$ day that could have been included in our sample, only three exhibit flares that lasted long enough to affect three long-cadence \emph{Kepler} data points, as we require for FLATW'RM.  The significant differences between our samples are likely to have led to our conflicting conclusions.  

Because the \citet{haw14,lur15,vid16,doy18} studies focused on stars that largely would not be included in our sample, we suggest that the correlation we see has potential to be appropriate for certain classes of stars. 

For the highest energy flares, the events investigated here seem to be observed regardless of their proximity to the starspot groups, including when the starspot group is out of view.  Because these flares are large, energetic events, it is possible that they can be seen not only when the flaring regions are located on the hemisphere facing \emph{Kepler}, but also over the limb.  The weaker flares, perhaps, follow the same trend, but are not strong enough to be seen over the stellar limb, appearing to occur more frequently around the starspot groups.  An additional explanation could be if some of the events are associated with a polar spot that can be seen regardless of the rotational phase of the object, and which would cause only small rotational variations in the light curve. 

In the case of the Sun, flares are observed mainly in bipolar regions between spots of different polarity. Solar flares, where reconnection occurs between different regions are rare. However, it is possible that on stars with stronger magnetic fields such reconnections between active nests are more frequent. This could explain why stronger flares seem to be less associated with the dark spots (as the measured timing of the flare is associated approximately with the middle of the brightening flare loop). On the Sun, in the case of bipolar spots, the leading spots are known to be larger and longer lived than the trailing sunspots \citep[see, e.g.,][]{mur14,zag15}. If this analog is true for other stars, too, that would mean that the weight of a detected active region is closer to the leading spot. This could explain the connection between the weaker flares and the starspots and the slight shift to $-0.1$ in phase instead of 0 in Figure \ref{hist}, suggesting that the size of the active region is on the scale of 0.2 phase \citep[$30^\circ$ diameter, which is not unlikely in photometric spot models, cf.][]{qiu17}. Furthermore, strong flares covering large area of the stellar disk can also wash away the exact timing of the event (on the Sun flare area is roughly connected to their strength).  For a simple model illustrating such an effect, see Appendix B.

\subsection{Caveats} \label{sec:caveats}

A number of complications in the data could move the location of the spot minima causing some of the phase differences to be shifted slightly.  In Section \ref{sec:analysis}, we accounted for a number of situations in which the spot minima could be altered.  There are, however, a few situations that could continue to be problematic and are not easily accounted for with the SAUCES algorithms.  We discuss those issues here.

\begin{itemize}

\item In the case where the adjacent rotations must be considered because the flare occurs near the starspot minimum, if evolution changes the starspot group significantly from one rotation to the next, using the adjacent rotations to estimate the actual time of minimum near the flare could lead to an inaccurate phase difference, $\Delta \varphi$.

\item Low-amplitude spots across the stellar surface or a polar spot will lead to undetectable changes in the light curve, but will bring active regions into and out of view and the detected starspot minima will be wrongly associated with the flares.  

\item A number of widely-separated spots rotating in and out of view could present in the light curve as a single spot structure \citep[e.g., Figures 3 and 7 of][]{roe17b}, which would shift the minimum associated with the starspot group.

\end{itemize}

\section{Conclusions} \label{sec:conclusions}

The sample of stars included here are at the cross-section of the \citet{mcq14} and \citet{dav16} samples, while also fitting our criterion of showing starspot evolution.  Our sample of 119 main-sequence stars with evolving starspots were run through the FLATW'RM algorithm \citep{vid18} to detect the flares and the SAUCES pipeline to identify the starspot minima and compare the timing of the flares and starspots.  

We find that there are 2447 flares that increase the observed flux at least $1\%$ over the star in 111 of the four-year, long-cadence \emph{Kepler} light curves in our sample.  Of these flares, 475 increase the flux of the star by at least $5\%$.  We found that though the flares were not  uniformly distributed across the surface, the smaller energy flares appeared to occur predominantly when the large starspot group was in view.  This could be a result of stronger flares being visible both when the flaring region is visible to \emph{Kepler} and when the region has crossed over the limb, but weaker flares are not strong enough to be seen over the limb, appearing to be more correlated with the starspots.  Another potential explanation could be the presence of polar starspots that are not detectable with \emph{Kepler} photometry alone.

To further investigate whether the starspot groups and the flares are connected, in future work, we will apply FLATW'RM and SAUCES to a wider variety of stars in the \emph{Kepler} archive.  However, efforts are required to further automate isolating the stars with evolving stellar activity.

\section*{Acknowledgements}

The authors thank the useful discussion with L.~van Driel--Gesztelyi, A.N.~Aarnio,  J.D.~Monnier, and A.B.~Davis. The authors acknowledge the Hungarian National Research, Development and Innovation Office grants OTKA K-109276, OTKA K-113117, and support through the Lend\"ulet-2012 Program (LP2012-31) of the Hungarian Academy of Sciences. KV is supported by the Bolyai J\'anos Research Scholarship of the Hungarian Academy of Sciences. This work has used K2 data from the proposal number GO12046. Funding for the \emph{Kepler} and \emph{K2} missions is provided by the NASA Science Mission directorate. 
This research has made use of the SIMBAD database,
operated at CDS, Strasbourg, France.

\facility{Kepler}

\software{FLATW'RM \citep{vid18}, SAUCES}

\section*{Appendix} \label{sec:appendix}

\subsection*{Appendix A} \label{sec:appendixA}

In order to test the abilities of SAUCES, we performed a series to tests using 1000-day-long model light curves.  For each of five model light curves, we included two starspots on a star with a rotation period of $P_\mathrm{rot} = 3.00$ days.  The larger spot was centered on a longitude of $120^\circ$ (0.33 in phase) and the smaller spot was centered on a longitude of $320^\circ$ (0.89 in phase).  The spots did not change over time in shape, size, darkness, or location (i.e., the spots do not evolve or show differential rotation with respect to one another).  While the stars in our observational sample do exhibit starspot evolution, we select our observational sample to be the instances of only a single starspot group.  Because we look at individual rotations, we do not see the effects of differential rotation.  Even in the cases of looking to the rotations before and after to better identify the starspot minimum, differential rotation is not dramatic in these systems.

For each of our four models, we included 50 flares distributed (A) from a uniform distribution, (B) all exactly at longitude $120^\circ$ (phase of 0.33), (C) from a uniform distribution within $\pm 0.05$ of phase 0.33, and (D) from a uniform distribution within $\pm 0.10$ of phase 0.33.  The flares are given random strengths with lifetimes generated from the flare strengths, following the description of a single-peak flare of \citet{dav14}. 
 
FLATW'RM recovers 47, 45, 48, and 49 flares, respectively.  The deviation from the 50 injected flares is the result of some smaller flares not satisfying the search criteria of FLATW'RM and SAUCES ignoring flares that occur less than one rotation within the beginning or the end of the light curve, cases which make identifying the nearest minima difficult in observational light curves. 

\begin{figure*}
\centering
\begin{subfigure}
\centering
\includegraphics[scale=0.35,angle=90]{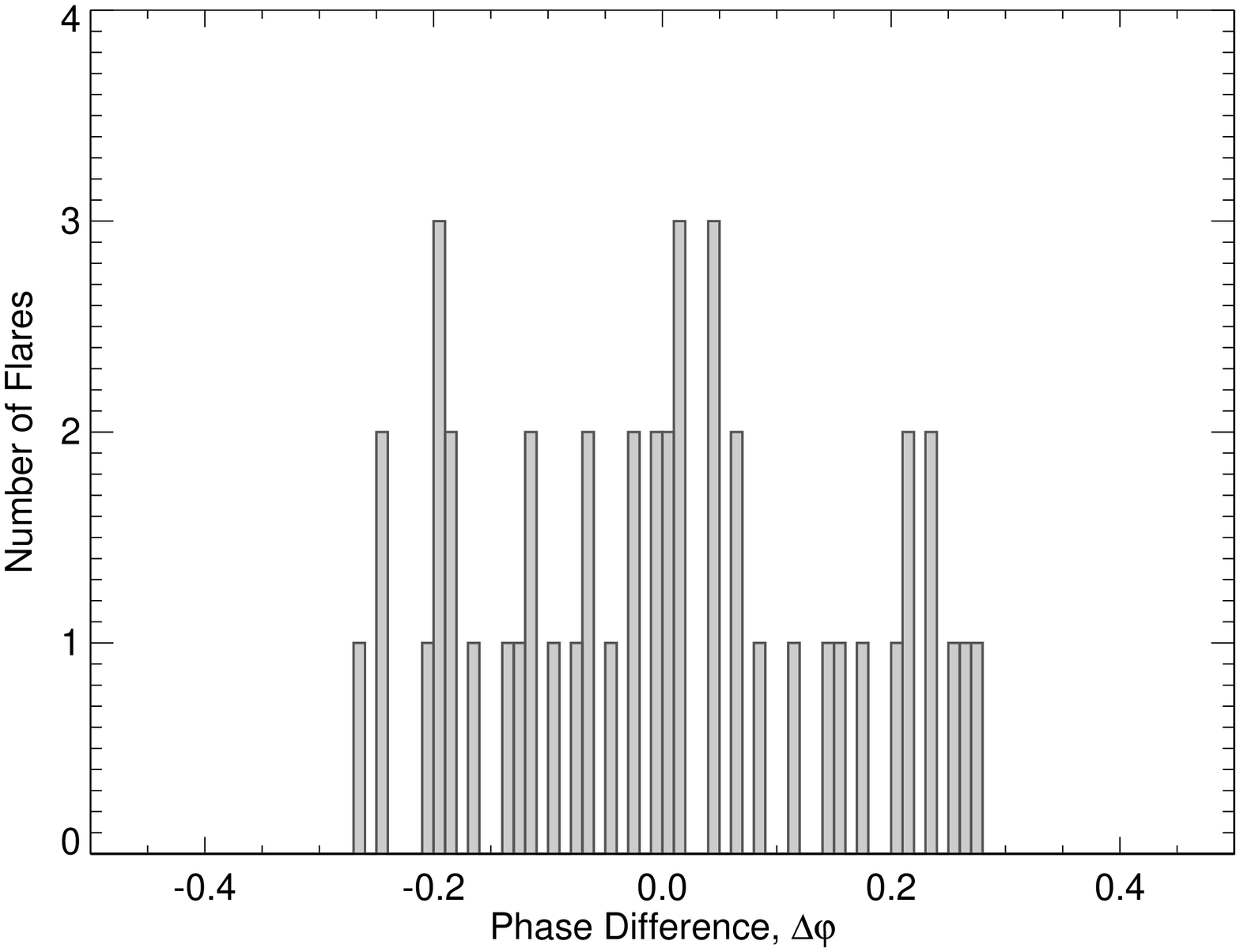}
\end{subfigure}
\begin{subfigure}
\centering
\includegraphics[scale=0.35,angle=90]{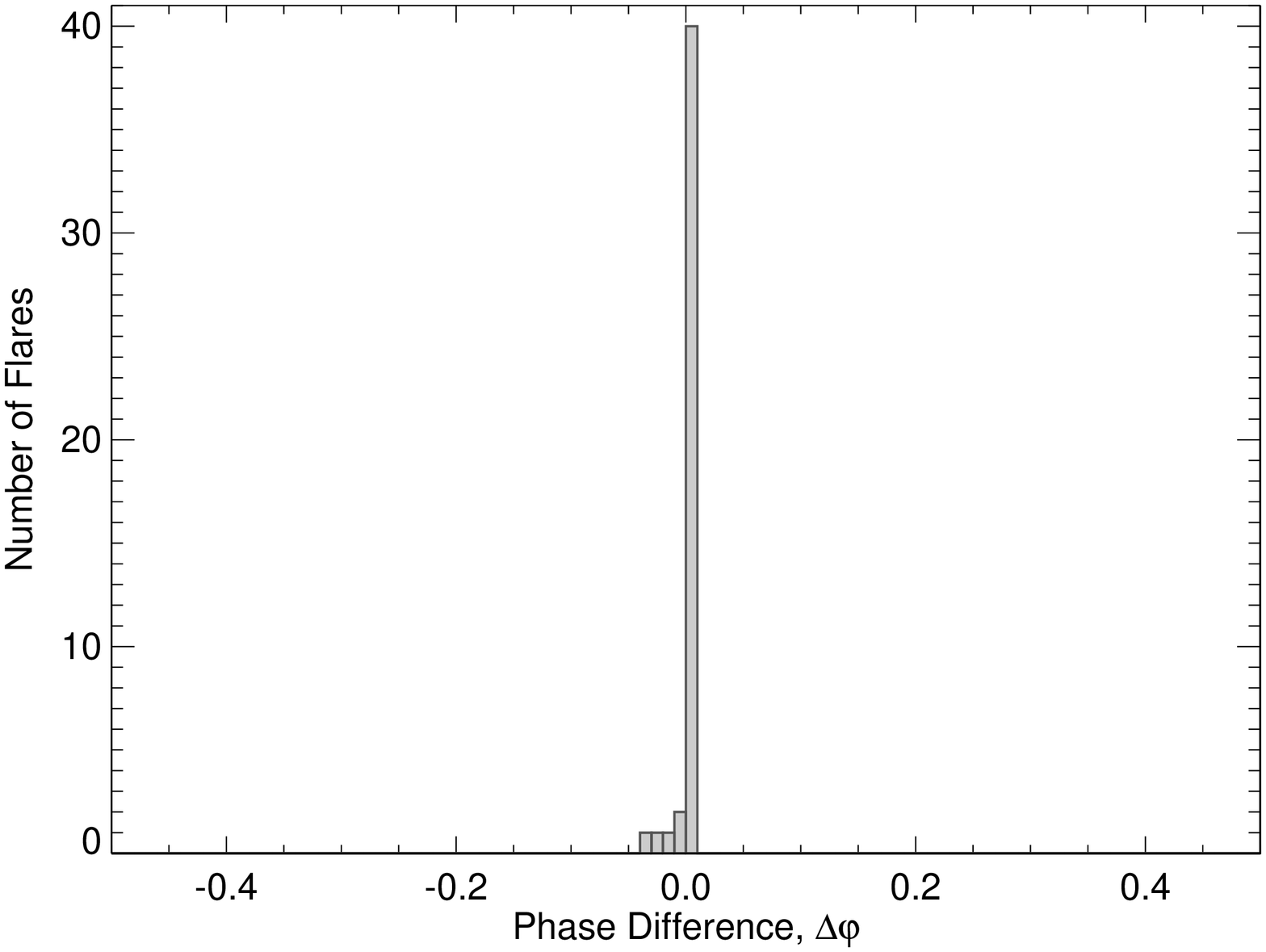}
\end{subfigure}
\begin{subfigure}
\centering
\includegraphics[scale=0.35,angle=90]{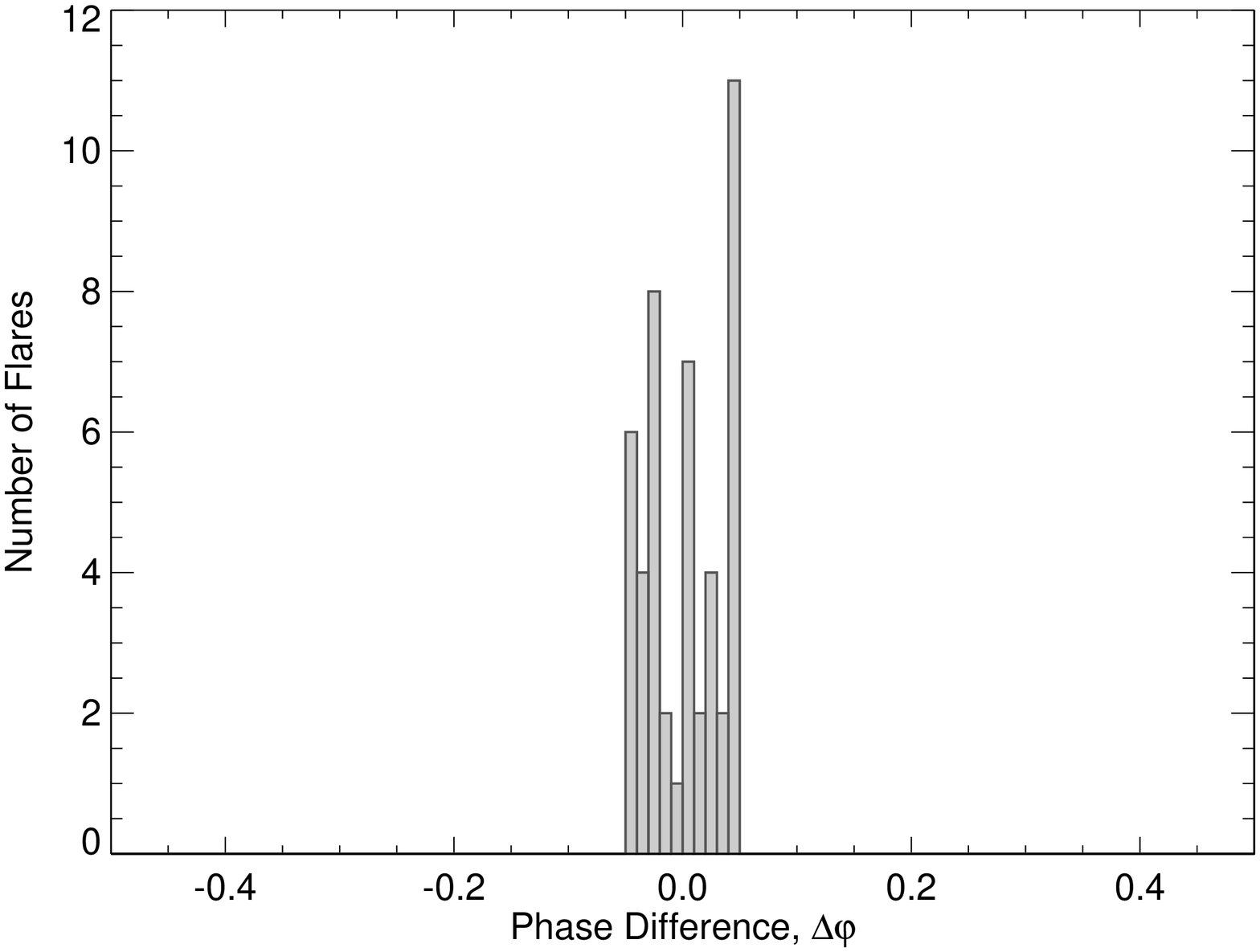}
\end{subfigure}
\begin{subfigure}
\centering
\includegraphics[scale=0.35,angle=90]{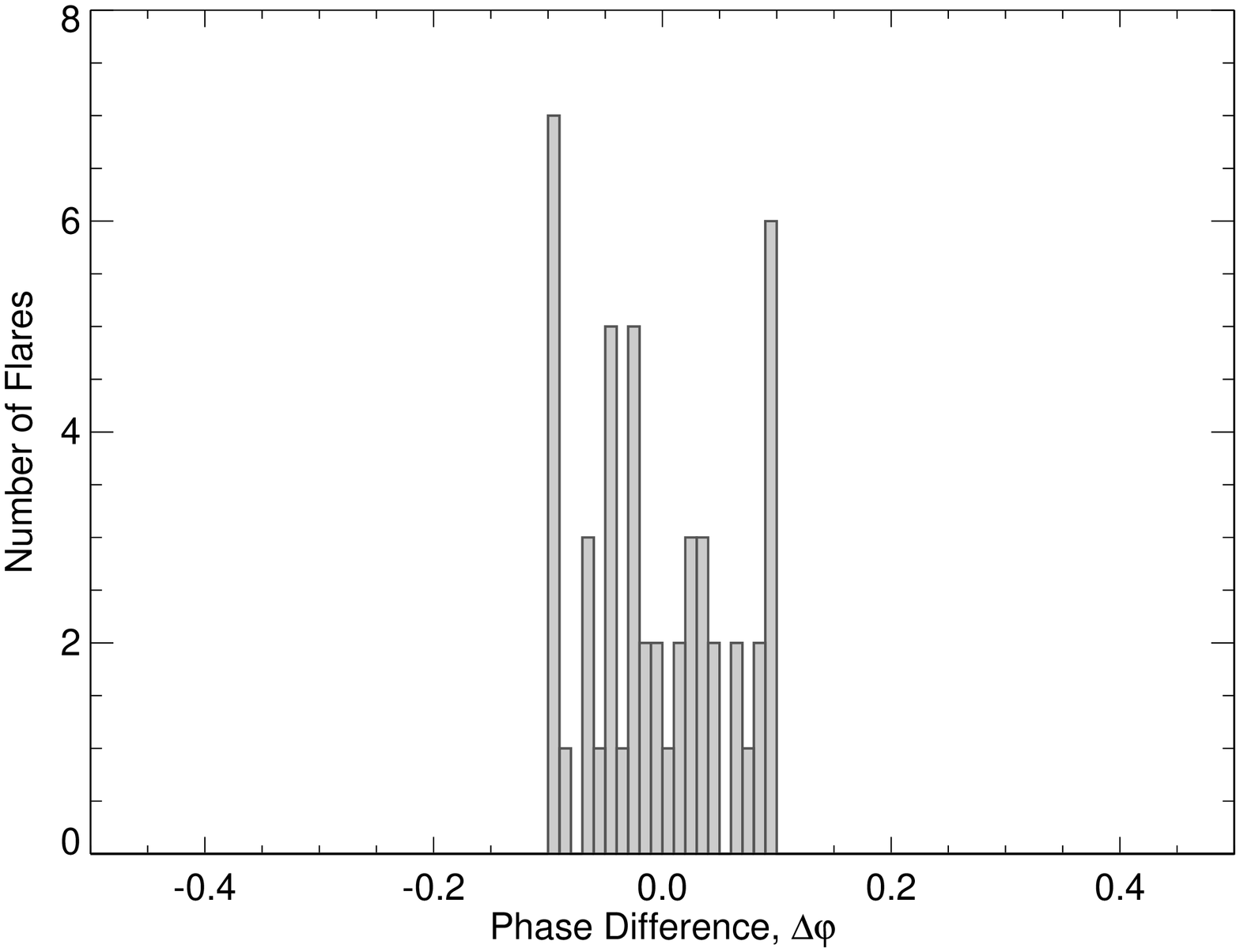}
\end{subfigure}
\caption{Histograms of the comparison of the phase difference between each model flare and the nearest model starspot.  Upper left: For Test A, the flares occurred over a uniform distribution of phases. Upper right:  For Test B, the flares occurred when the spot centered on longitude $120^\circ$ faces the observer (phase = 0.33).  Lower left:  For Test C, the flares occurred within $\pm 0.05$ in phase of the spot centered on longitude $120^\circ$ faces the observer (phase = 0.33).  Lower right: For Test D, the flares occurred within $\pm 0.10$ in phase of the spot centered on longitude $120^\circ$ faces the observer (phase = 0.33).  Each bin represents 0.01 in phase.}
\label{testsABCD}
\end{figure*}

For the user-supplied parameters of FLATW'RM, we required each flare contain at least three data points.  For more details on the FLATW'RM algorithm, see \citet{vid18}.  We apply the SAUCES algorithms to the test data sets as described in Section \ref{sec:analysis}.  We calculate the difference in phases, $\Delta \varphi$, between the starspot minima and the maximum brightness of the flares.  We show the results accounting for the effect of the flares on the surrounding light curve in Figure \ref{testsABCD}.

We note that for Tests B--D, none of the flares are close enough to the smaller starspot located at a longitude of $320^\circ$ to cause SAUCES to compare the timing of flare to that starspot.  Because the flares of Test A can occur anywhere in the light curve with respect to the starspots as a uniform distribution, it is not surprising that the smaller phase differences have more flares than the larger phase differences.  Unlike in the cases of single spots, the largest separation possible for Test A between a flare and a spot is $\pm 0.28$, which is observed (compared to $\pm 0.50$ when only one spot is present).  That phase difference is only possible when the starspot at longitude $120^\circ$ is rotating out of view before the other spot comes into view.  Phase differences of $-0.22 \le \Delta \varphi \le +0.22$ are accessible as either spot rotates out of view. Because there are two spots in the model light curve, SAUCES correctly finds phase differences that are only less than $\Delta \varphi \approx 0.28$, the maximum phase difference a flare can be from a local minimum.  Therefore, we do not expect a uniform distribution to be returned in our histogram.

We note that Test A differs from our observed sample in that there are two spots present, a case we do not consider in the observed sample, but include here for the illustrative purpose of the difficulty differentiating the nearest starspot when multiple are present.  While the second spot is still present for Tests B--D, the flares by definition do not occur close enough to the second spot for SAUCES to calculate the $\Delta \varphi$ between the flare and the smaller spot at longitude $320^\circ$.

In Test B, we see that in most cases, SAUCES does properly account for flares that occur in the center or the starspot signature.  The cases here where the detection is offset from $\Delta \varphi = 0$ are related to flares occurring in at rotations such that SAUCES cannot properly determine the minimum (e.g., at least three sequential rotations or three of four sequential rotations), affecting SAUCES' ability to accurately locate the minimum (see Section \ref{sec:analysis} for details on how the algorithm handles flares in local minima).  Tests C and D reproduce the distributions which were injected, and are also subjected to the same potential problems as the other tests.

While the location of some flares may be skewed when flares happen in sequential minima, we found that this was the case for only $1.8\%$ of the flares in our observational sample.  The effect is greater here in Test B because the flares are limited to only one phase and must then occur in $15\%$ of the rotations.

\subsection*{Appendix B} \label{sec:appendixB}

Similarly to the study in Appendix A, here we model a 1000-day long light curve with $P_\mathrm{rot} = 3.00$ days.  Two spots are again present, but they are now located adjacent to each other at the same latitude.  The ``leading'' spot, or the first spot to appear over the limb as the star rotates, is defined to have a radius of $20^\circ$ on the surface of the star.  The ``trailing'' spot is assigned a radius of $15^\circ$ on the surface of the star.  The proximity of the spots and the difference in their sizes creates a light curve shape where there is only one local minimum, when the center of the larger spot is facing the observer.  

With this model, we aim to investigate the potential of a starspot group with a leading-trailing pair of starspots (like that described in Section \ref{sec:discussion}).  The leading spot is slightly larger, and the location at which the flare occurs varies depending on the test.  

In Test E, the flares are uniformly distributed across the light curve.  In Test F, the flares occur only when the point between the starspots is facing the observer.  In Tests G and H, the flares occur within $\pm 0.05$ and $\pm 0.10$ in phase of when the point between the starspots is facing the observer.  These phase difference limits are chosen as the flares occurring within this range will still be strong associated with the starspots (the sum of the starspots' radii spans approximately 0.1 in phase).  With Tests F--H, the over-density of flares in our observed sample near $\Delta \varphi = -0.1$ is recreated (at a different phase, dependent upon the spot characteristics), suggesting that such a scenario of starspot groups in an active nest is possible, as described in Section \ref{sec:discussion}.

\begin{figure*}
\centering
\begin{subfigure}
\centering
\includegraphics[scale=0.35,angle=90]{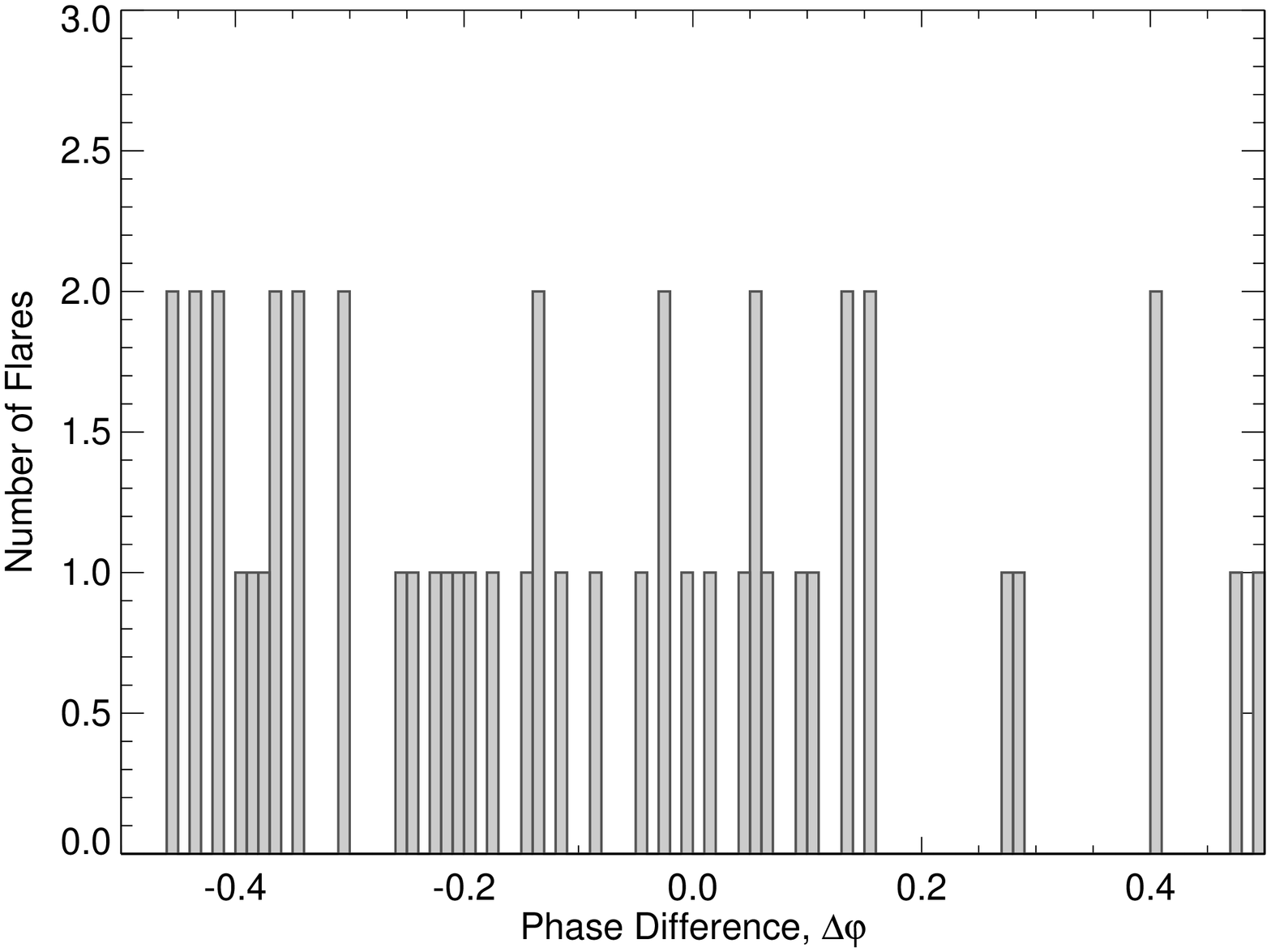}
\end{subfigure}
\begin{subfigure}
\centering
\includegraphics[scale=0.35,angle=90]{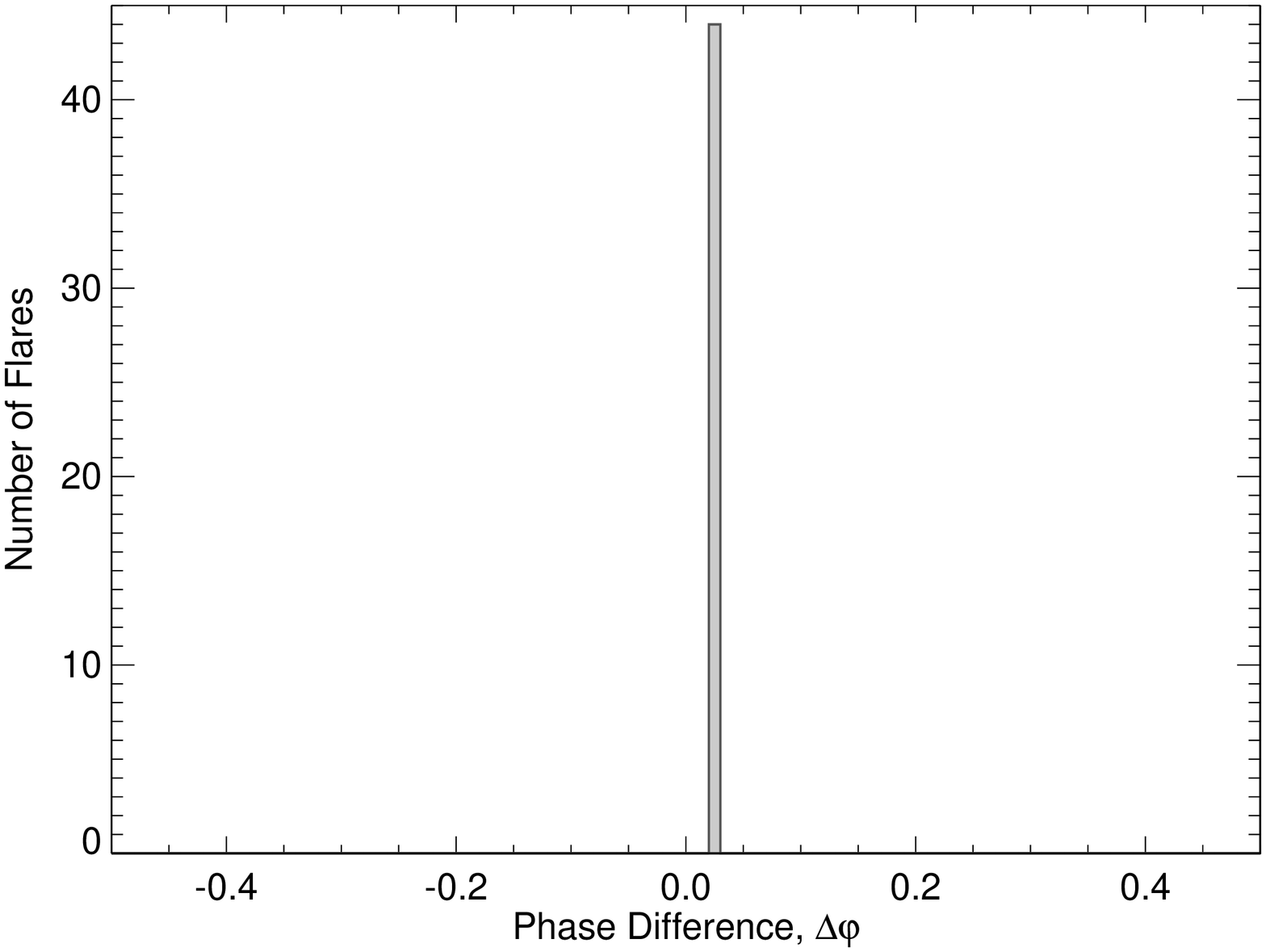}
\end{subfigure}
\begin{subfigure}
\centering
\includegraphics[scale=0.35,angle=90]{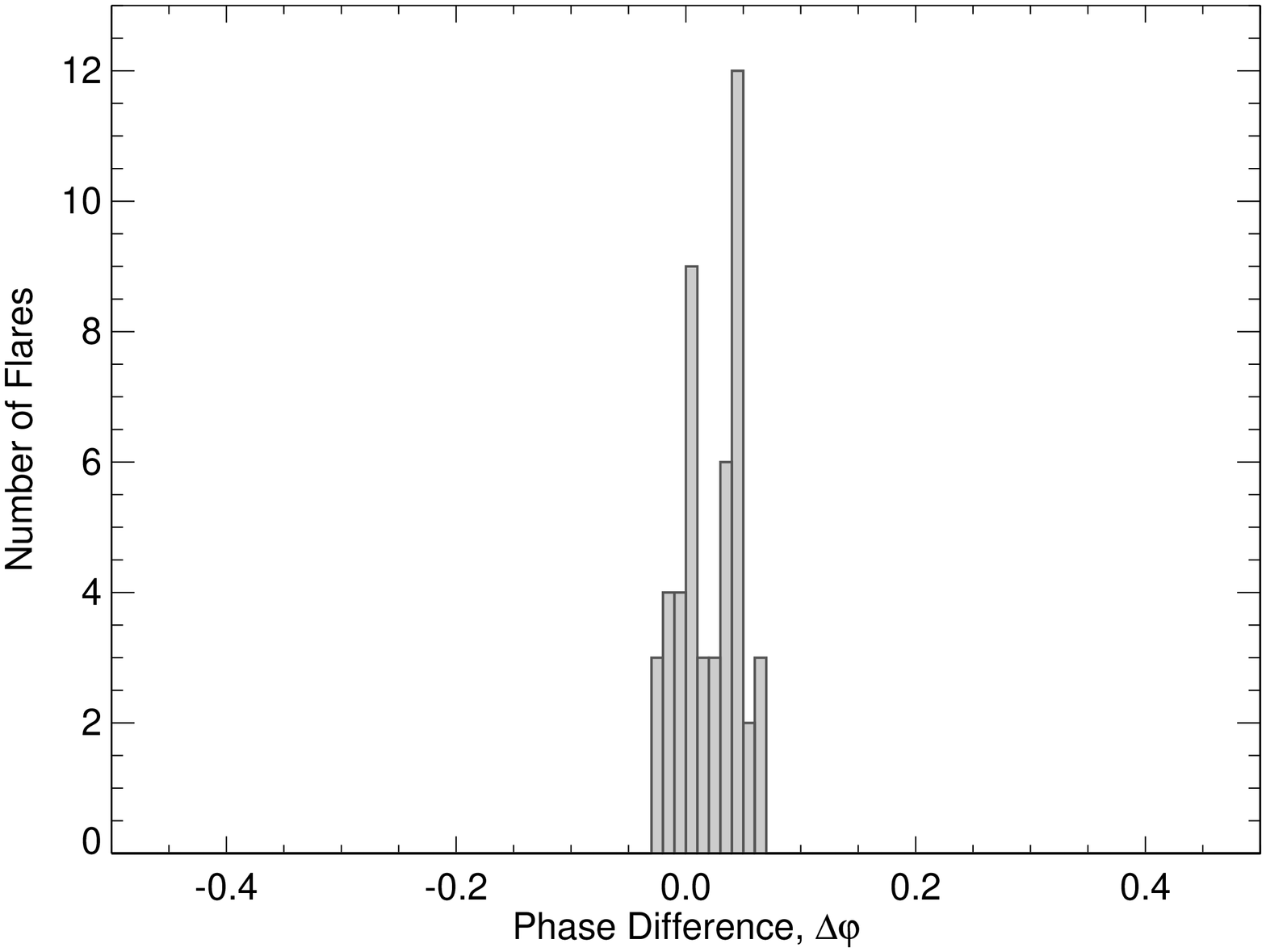}
\end{subfigure}
\begin{subfigure}
\centering
\includegraphics[scale=0.35,angle=90]{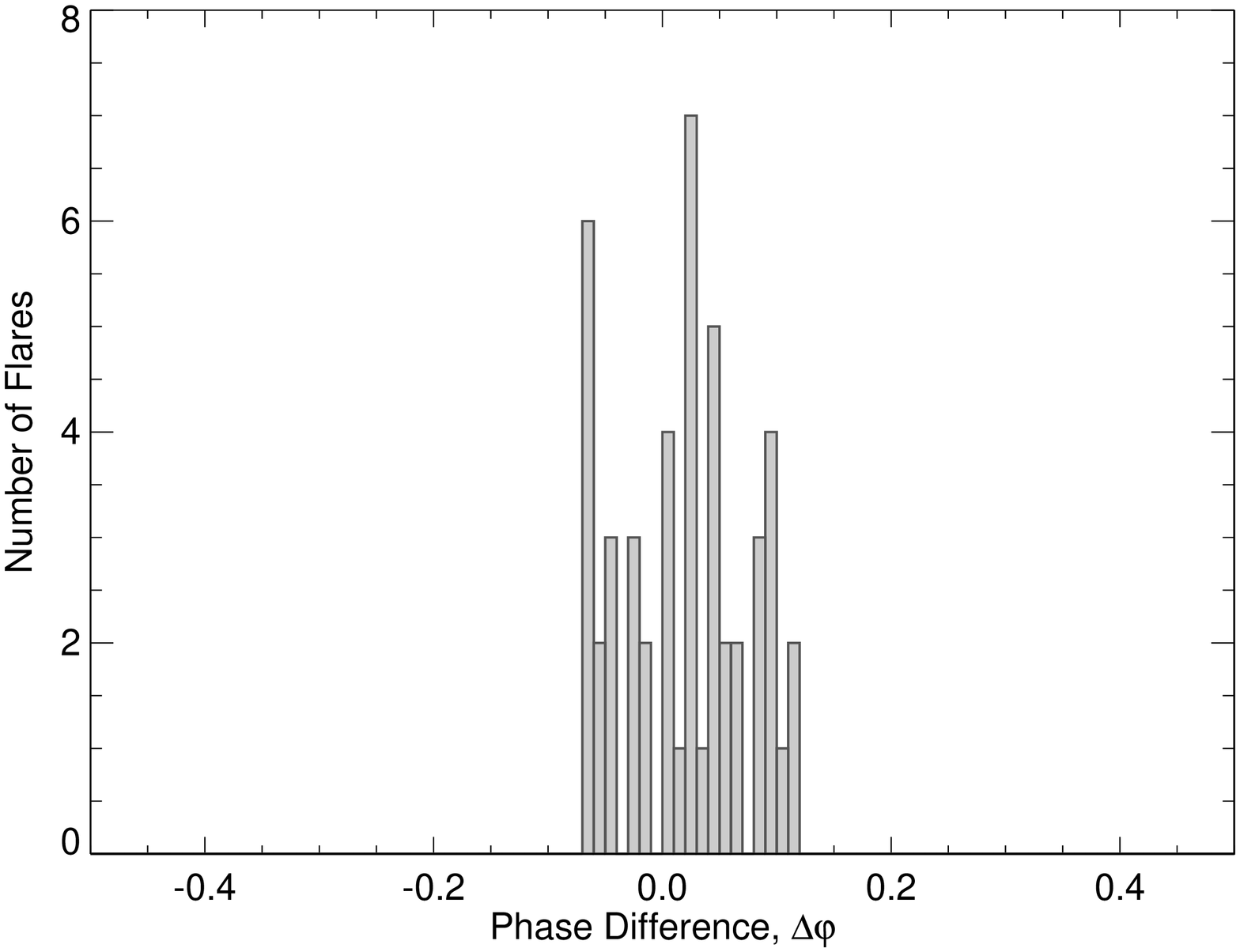}
\end{subfigure}
\caption{Histograms of the comparison of the phase difference between each model flare and the nearest model starspot for a leading-trailing spot configuration.  Upper left: For Test E, the flares occurred over a uniform distribution of phases. Upper right:  For Test F, the flares occurred when the point between the starspots is facing the observer.  Lower left:  For Test G, the flares occurred within $\pm 0.05$ in phase of the point when the point between the starspots is facing the observer.  Lower right: For Test H, the flares occurred within $\pm 0.10$ in phase of the point when the point between the starspots is facing the observer.  Each bin represents 0.01 in phase.}  
\label{testsEFGH}
\end{figure*}

Each light curve included 50 flares, and 48, 44, 49, and 48 were recovered by FLATW'RM.  Again, this deviation from 50 flares is the same as that in Appendex A, where the flares can go undetected if they are weak, of short duration, or occur too close to the beginning or end of the light curve.  The results of applying the SAUCES alogrithms are shown in Figure \ref{testsEFGH}.

While no single star in our sample exhibits the features displayed in these tests, the combination of all stars in the sample does show a correlation between the starspots and the flares with flux increases between $1\% - 5\%$.  The correlation, as well as the particular over-density near $\Delta \varphi = -0.1$, may be explained by many stars exhibiting behavior similar to that shown in Tests F--H.

\subsection*{Appendix C} \label{sec:appendixC}

The stars in our sample range in spectral type from late-F to early-M, according to masses given by \citet{dav16}.   Here, we cut the distribution in mass at $0.78 \ M_\odot$, approximately the mass boundary between G and K stars.  Our objective was to determine whether or not the distinct distribution present in Figure \ref{hist} would be associated with only certain masses of main sequence stars.  

\begin{figure*}
\centering
\begin{subfigure}
\centering
\includegraphics[scale=0.35,angle=90]{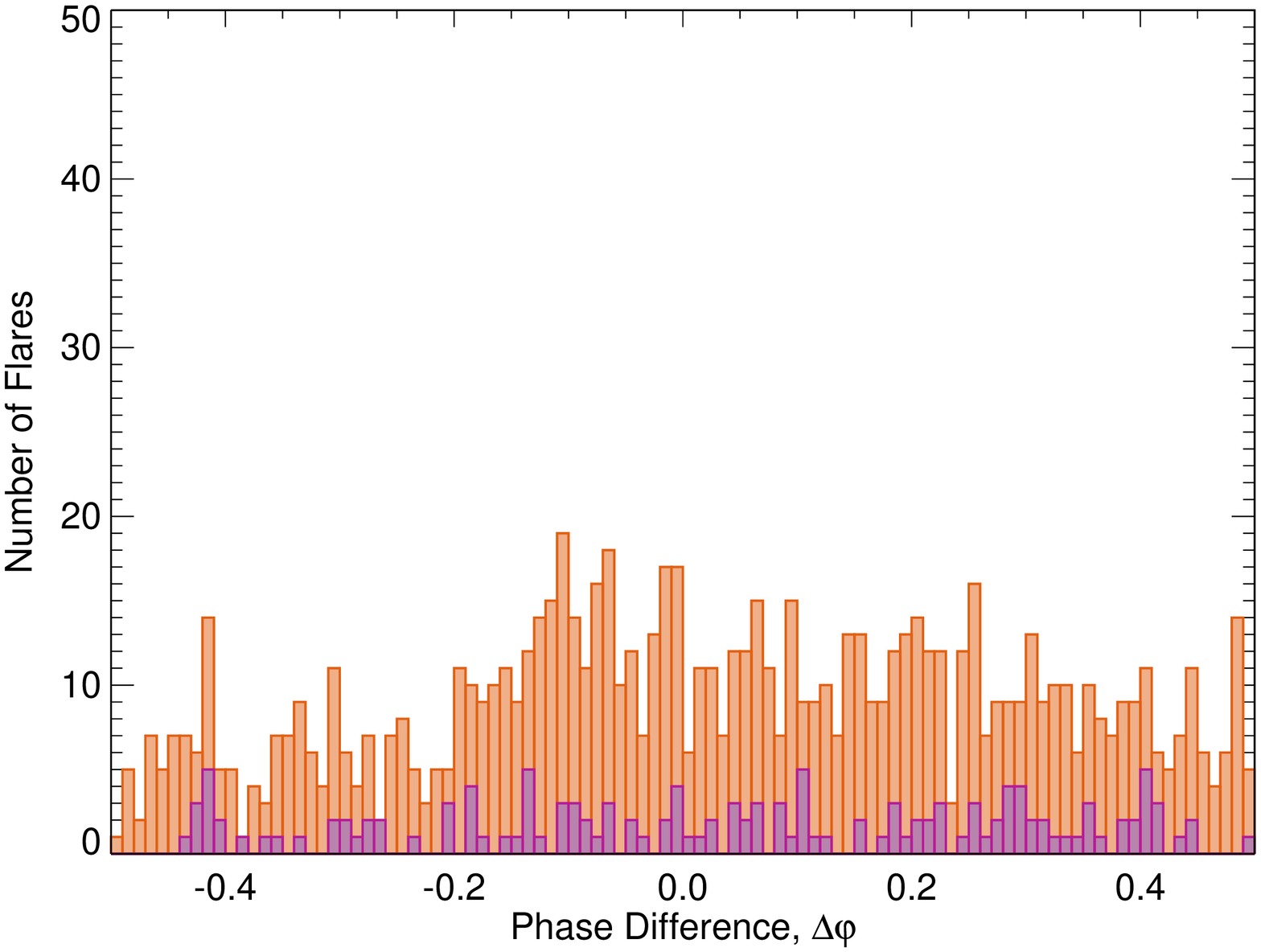}
\end{subfigure}
\begin{subfigure}
\centering
\includegraphics[scale=0.35,angle=90]{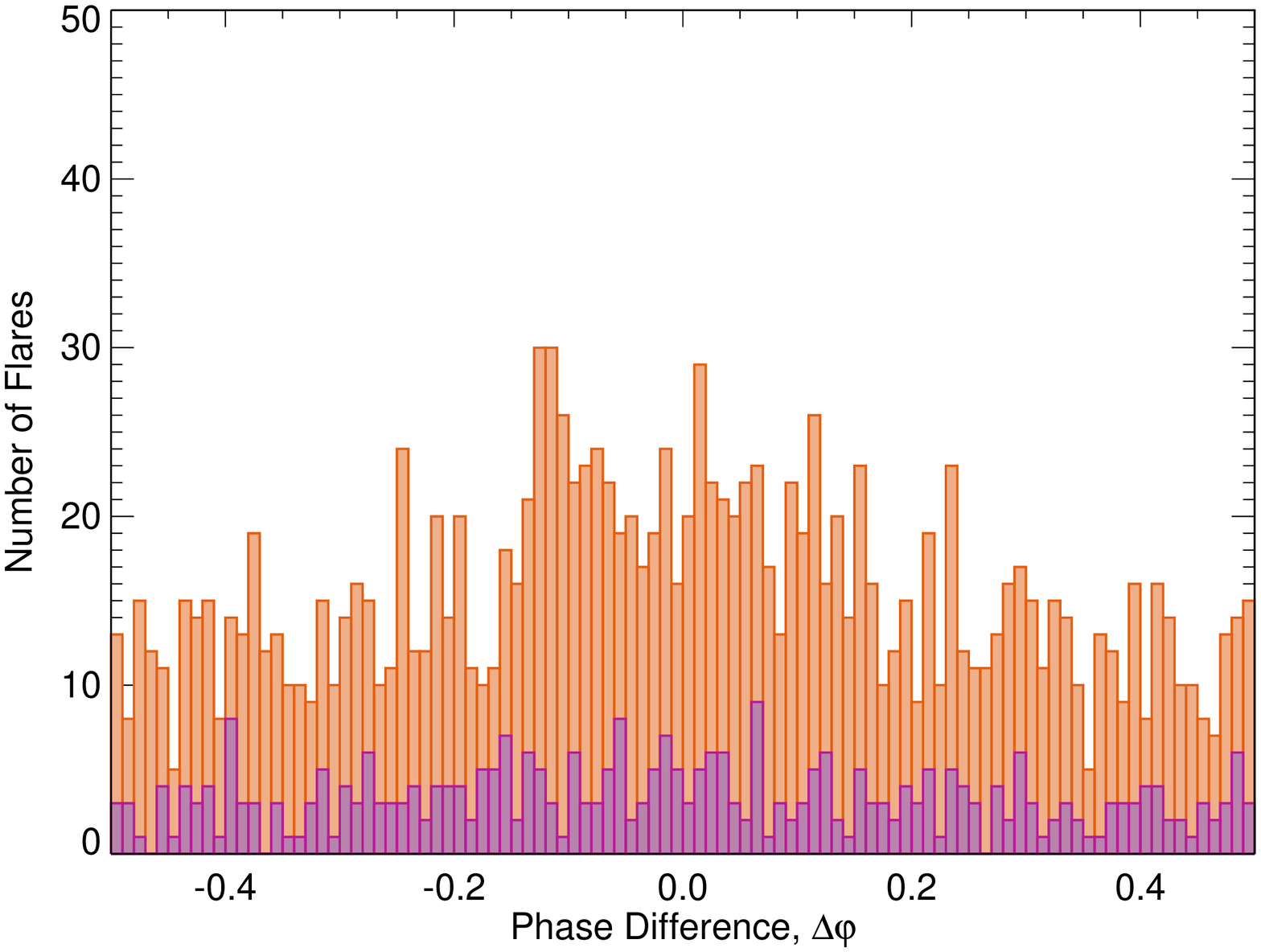}
\end{subfigure}
\caption{Histogram of the number of flares occurring in bins of phase difference $\Delta \varphi = 0.01$.  The color scheme follows that of Figure \ref{hist}. Left:  Flares occurring on the late-F and G-type stars in the sample.  The increased number of flares in the phases near the starspots is weakly evident.  Right:  Flares occurring on the K and early-M-type stars in the ample.  The lower-energy (orange) flares are more numerous in the phases closely surrounding the starspots. }
\label{FGandKM}
\end{figure*}

In Figure \ref{FGandKM}, the histograms for F/G stars (left) and K/M stars (right) are shown to both have a concentration of flares occurring near the starspots for flares with flux increases between $1\% - 5\%$.  While the peak is stronger for the late-type stars, more flares occurred in this sample of light curves, potentially strengthening the single.  Therefore, the phenomenon of these weaker flares being associated with the starspots is observed in all spectral types of convective-envelope, main-sequence stars.

\bibliography{starspotpapers}

\end{document}